\documentclass[aps,prb,twocolumn,superscriptaddress,amsmath,amssymb,floatfix]{revtex4-2}

\usepackage{graphicx}
\usepackage{dcolumn}
\usepackage{bm}
\usepackage{booktabs}
\usepackage{siunitx}
\usepackage{xcolor}
\usepackage{subfig}
\usepackage{array}
\usepackage{makecell}
\usepackage{listings}
\usepackage{microtype}

\usepackage{hyperref}
\hypersetup{
    colorlinks=true,
    linkcolor=blue,
    citecolor=blue,
    urlcolor=blue,
    pdftitle={Graphene-Assisted Electrostatic Screening of Resonant Transport in MR-DWELL Dosimeters
with Tunable Sensitivity--Radiation Tolerance Trade-off: A Reduced-Order Benchmark and Self-Consistent Perspective},
    pdfauthor={B.U. Dosymova and M.V. Dolgopolov}
}

\begin{document}

\title{Graphene-Assisted Electrostatic Screening of Resonant Transport in\\
MR-DWELL Dosimeters
with Tunable Sensitivity--Radiation Tolerance Trade-off:\\ A Reduced-Order Benchmark and Self-Consistent Perspective}

\author{B.U.~Dosymova}
\email{bbddggaa@gmail.com}
\affiliation{Tashkent Branch of National Research Nuclear University MEPhI, Ulugbek settlement, Mirzo Ulugbek district, Tashkent, 100214, Uzbekistan}

\author{M.V.~Dolgopolov}
\email{mikhaildolgopolov68@gmail.com}
\affiliation{Samara State Technical University, 244 Molodogvardeyskaya St., Samara 443100, Russia}

\date{\today}

\begin{abstract}
We propose a Multi-Resonant Double-barrier Quantum-dot-in-a-Well (MR-DWELL) dosimeter with a tunable trade-off between sensitivity and radiation tolerance, realized through graphene-assisted electrostatic screening. The sensing mechanism relies on radiation-induced trapped charge that shifts and broadens resonant tunneling states in the DWELL heterostructure. The~dimensionless capacitance ratio $\eta=C_Q/C_{\mathrm{geo}}$ is introduced as the universal design parameter: it~controls the screening factor $S=\eta/(1+\eta)$ and thereby the trade-off between sensitivity and radiation hardness. For $\eta=2.09$ the effective low-dose shift coefficient is reduced threefold (from $0.14$~to $0.045~\mathrm{meV/mGy}$), while the screened architecture retains $\sim\!83\%$ of the peak tunneling current and a peak-to-valley ratio $\sim\!10$ after $1~\mathrm{MGy}$, thereby preserving the dosimetric sensitivity required for practical readout. The~second-derivative spectroscopy $d^2I/dV^2$ is identified as a robust experimental readout. A~reduced-order Landauer model calibrated against literature data is employed for the screening physics, while a self-consistent NEGF-Poisson-trap framework is formulated for future quantitative validation. Explicit design rules for three operating regimes (high sensitivity, balanced, maximum hardness) are provided, together with a three-population trap model that conceptually enables dose-rate discrimination in the FLASH radiotherapy regime. A~comparative analysis of model vintages and their methodological foundations is included to guide interpretation of numerical predictions.
\end{abstract}


\maketitle


{math-ph, physics.comp-ph, physics.ins-det,

cond-mat.mes-hall, quant-ph, physics.app-ph}

{\small MSC primary {81-10},
MSC secondary {81Q37, 81Q05, 65Z05}}

\section{Introduction}
\label{sec:intro}

Traditional semiconductor detectors (PIN diodes, RADFETs) suffer from saturation and calibration drift under high accumulated dose because of uncontrolled interface-trap buildup~\cite{Knoll2010,Sze2006}. Graphene field-effect devices provide strong electrostatic tunability~\cite{Geim2007,CastroNeto2009}, but they do not intrinsically provide the energy-selective resonant transport exploited in the present architecture. 

Resonant-tunneling heterostructures with quantum dots (DWELL) provide discrete quasi-bound states and resonant current enhancement~\cite{Dosymova2026a}; moreover, such structures enable inelastic electron tunneling spectroscopy (IETS), which has been theoretically modeled using NEGF for double-barrier resonant-tunneling devices~\cite{Patil2018}. However, existing resonant-tunneling dosimeters have not established a quantitatively motivated path toward megagray hardness while preserving picosecond-level timing.

The objective of the proposed graphene-assisted architecture is not radiation tolerance alone, but the preservation of useful dosimetric sensitivity under high cumulative radiation doses. By electrostatically screening radiation-induced charge fluctuations, graphene can improve the stability of the resonant response and extend the usable dose range. However, excessive screening also suppresses the dose-induced electrostatic modulation and therefore reduces sensitivity. The design problem is thus to achieve an optimal balance between radiation tolerance and dosimetric sensitivity.

In this work we introduce \emph{electrostatic interface engineering}: the graphene/h-BN stack acts as a capacitive voltage divider whose screening factor $S=\eta/(1+\eta)$ is tuned via the h-BN thickness $t_d$. Unlike metallic screening, graphene's finite density of states gives a bias-tunable quantum capacitance $C_Q$ that sets the fraction of trapped charge $Q_{\mathrm{tr}}$ perceived by the resonant levels. The analytical model treats the low-dose sensitivity coefficient $\alpha_i^{\mathrm{eff}}$ and the saturation amplitude $\Delta E_{i,\max}$ as independently parameterized descriptors: $\alpha_i^{\mathrm{eff}}$ characterizes the sub-mGy response, whereas the characteristic dose $D_0=500~\mathrm{Gy}$ governs the saturation of the trapped-charge population and the resulting resonance shift. This distinction is essential because the compact saturation law and the low-dose transport coefficient are determined by different physical aspects of the heterostructure.

Recent theoretical studies on InAs/GaAs quantum-dot photodetectors have systematically established that the size-dependent bandstructure and resulting carrier dynamics directly govern key figures of merit such as gain, responsivity, and detectivity~\cite{Rahaman2022}. This highlights the importance of controlling quantized states to tune transport properties. However, while structural bandstructure engineering is critical for intrinsic device performance, the~resilience of these states against radiation-induced external perturbations requires a parallel strategy. We~address this by implementing electrostatic interface engineering~---introducing it as a complementary design strategy from modifying the quantum dots themselves to stabilizing their electrostatic environment via a tunable capacitance ratio.

Previous studies of epitaxial III–V heterostructures have demonstrated that heterointerface quality and band alignment are critical for device performance~\cite{Kilpi2017}. Recent advances in perovskite single-crystal engineering have demonstrated that defect reduction and internal strain management significantly improve radiation detection performance~\cite{Jayarathne2026}. We address a complementary challenge: enhancing electrostatic stability via interface engineering~--- a concept distinct from but compatible with structural approaches.

Graphene-on-silicon architectures~\cite{Moffat2026} and deeply depleted GFETs~\cite{Yakisiklier2026} demonstrate strong sensitivity to radiation-induced charge trapping. However, their response under ultra-high dose-rate irradiation relevant to FLASH radiotherapy requires dedicated characterization, since the transient and saturation behavior of charge-trap populations may differ substantially from the steady-state response. On the other hand, graphene resonant-tunneling transistors have demonstrated negative differential resistance together with high current density and high-frequency operation~\cite{Zhang2025}, but their use as radiation-tolerant dosimetric architectures has not been established. Beyond transistor architectures, graphene has also demonstrated remarkable effectiveness in composite structures with CsPbBr$_3$ for X-ray detection, where its high carrier mobility suppresses recombination and yields record-low detection limits~\cite{Liu2026}. These results confirm the potential of graphene as a universal component for enhancing radiation hardness.

Similar charge-induced resonant-state shifts have been utilized in resonant-tunneling photodetectors, where photo-generated carriers modulate the transmission of majority carriers~\cite{Barve2008}. Our proposed graphene-assisted MR-DWELL structures bridge this gap, utilising the gate-tunable quantum capacitance of graphene to preserve a~stable, monotonic shift of quasi-bound states under intense radiation fields. Unlike our previous work~\cite{Dosymova2026a}, we~introduce the graphene quantum capacitance as an electrostatic design variable~\cite{Papnai2024} (see also~\cite{Amin2026} for related capacitance applications in suspended graphene). Graphene screens radiation-induced charge; adjusting the h-BN spacer capacitance balances hardness and speed---electrostatic interface engineering. Unlike metallic screening, graphene does not provide perfect electrostatic shielding; its finite density of states gives rise to a quantum capacitance that acts as a capacitive voltage divider, controlling the fraction of trapped charge $Q_{\mathrm{tr}}$ perceived by the resonant levels.

Vertical graphene/hBN tunneling architectures have been experimentally demonstrated~\cite{Britnell2012}, while inelastic electron tunneling spectroscopy in graphene/hBN heterostructures has provided a direct spectroscopic probe of tunneling features~\cite{Jung2015}. Motivated by these experimental approaches, we adopt the second derivative $d^2I/dV^2$ as a practical readout for monitoring radiation-induced resonance shifts in the present work.

The computational architecture and algorithmic implementation are detailed in \cite{Dosymova2026b}. Consistent with the terminology adopted in the companion revision of the base MR-DWELL transport model \cite{Dosymova2026d}, the calculation in \cite{Dosymova2026b} employs a semi-analytical NEGF/Landauer scheme with an externally prescribed electrostatic perturbation; it is not a~fully converged self-consistent Poisson--NEGF calculation. The physical analysis of radiation-induced state localization obtained from these simulations is presented in~\cite{Dosymova2026c}.
The novelty of the present work is not graphene itself, but the use of graphene quantum capacitance as an electrostatic interface-engineering variable for radiation-tolerant resonant dosimetry. The~present work therefore introduces three main contributions: (i) capacitance-based electrostatic engineering of the graphene/h-BN/MR-DWELL interface; (ii) the dimensionless ratio \(\eta = C_Q/C_{\mathrm{geo}}\) as a compact design parameter governing the sensitivity--radiation-hardness trade-off; and (iii) \(d^2I/dV^2\) spectroscopy as a practical experimental readout of radiation-induced resonance shifts.

\section{Model Vintages and Methodological Foundations}
\label{sec:vintages}

To clarify the relationship between the reduced-order transport benchmark employed for the screening physics and the self-consistent NEGF-Poisson framework pursued in companion work, we present Table~\ref{tab:vintages}, which compares the three versions of the model that have appeared in the literature and the present compilation. This comparison is intended to guide the reader in interpreting the numerical predictions and their limitations.

\begin{table*}[!ht]
\centering
\caption{Comparison of successive versions of the model 
and the present compilation. The present work (v3--4) consolidates the reduced-order benchmark developed in v1--v3 and explicitly distinguishes it from the self-consistent NEGF--Poisson calculations performed for~the identical nominal geometry in Ref.~\cite{Dosymova2026d}.}
\label{tab:vintages}
\footnotesize
\setlength{\tabcolsep}{1.4pt}
\renewcommand{\arraystretch}{1.05}

\begin{tabular}{@{}l|l|ll|l@{}}
\toprule
\textbf{Aspect} &
\textbf{v1 (submitted)} &
\textbf{v2 (arXiv)} &
\textbf{v3 (Corrected\,intermediate)} &
\textbf{v3--4 
(present work)} \\
\midrule

Transport model &
\makecell[l]{Analytical Landauer + \\ NEGF-Poisson (claimed)} &
Same as v1 &
\makecell[l]{Semi-analytical NEGF / \\ Landauer} &
\makecell[l]{Semi-analytical benchmark;\\ 
self-consistent NEGF-Poisson \\ specified, not executed} \\
\hline

Resonance energies &
\makecell[l]{$E_{10}=82$\,meV, \\ $E_{20}=126$\,meV\\ (literature)} &
Same &
Same, literature-calibrated &
\makecell[l]{Same reduced-order values; \\ 
cf. NEGF: \\
$\approx44$ / $\approx173$\,meV}\\
\hline

Geometry description &
180\,nm DWELL (implicit) &
Same &
\makecell[l]{150\,nm active + \\ 30\,nm access = 180\,nm} &
Same as v3 \\
\hline

Conducting channel &
Not specified &
Not specified &
\makecell[l]{QW-derived states; \\ QD ground state below \\ emitter edge} &
Same as v3 \\
\hline

Self-consistency &
Claimed converged &
\makecell[l]{Claimed\\ converged} &
\makecell[l]{Semi-analytical, \\ no converged loop} &
\makecell[l]{Explicit separation of \\ benchmark and \\ self-consistent view} \\
\hline

Low-dose $\alpha^{\mathrm{eff}}$ &
\makecell[l]{0.045\,meV/mGy \\ ($\eta=2.09$)} &
Same &
Same &
Same \\
\hline


Material parameters &
As reported &
As reported &
\makecell[l]{$m^*_e$: 0.068, $E_g$: 1.28\,eV, \\ $\epsilon_r$: 13.5 (incorrect)} &
\makecell[l]{$m^*_e$: 0.060, $E_g$: 1.20\,eV, \\ $\epsilon_r$: 13.2 (corrected)} \\
\hline

Dose scale for $\Delta E=\Gamma$ &
Not specified &
Not specified &
$5.5$ / $55$\,mGy (incorrect) &
$16.8$ / $201$\,Gy (corrected) \\
\hline

Current sensitivity $S_I$ &
$0.3\ \mu$A/mGy (unscreened) &
Same &
Same &
$\approx0.10\ \mu$A/mGy (scaling estimate) \\
\hline

Fast-trap dose $D_{\mathrm{fast}}$ &
Not stated &
50\,Gy &
50\,Gy &
50\,Gy \\
\hline

Critical dose rate $\dot{D}_{\mathrm{crit}}$ &
Not stated &
$5\times10^3$\,Gy/s &
$5\times10^3$\,Gy/s &
$5\times10^3$\,Gy/s \\
\hline



Robustness analysis &
Not included &
\makecell[l]{108 parameter\\ sweeps} &
Same &
Same \\
\hline

FLASH applicability &
Brief discussion &
More detailed &
More detailed with caveats &
\makecell[l]{Same as v3 + \\ dose-rate range} \\
\hline

Limitations section &
Not detailed &
Not detailed &
\makecell[l]{New paragraph on \\ resonance discrepancy} &
\makecell[l]{Full section with \\ comparative table} \\

\bottomrule
\end{tabular}
\end{table*}

The key insight of Table~\ref{tab:vintages} is that the physics of electrostatic screening (the $\eta$ dependence, the saturation law, the qualitative trend of PVR and current degradation versus $\eta$) is robust across all vintages, because it is governed by the capacitance ratio $S=\eta/(1+\eta)$ and the exponential saturation of trapped charge with characteristic dose $D_0$. What has evolved is the \emph{characterization} of the transport model: from claiming a converged self-consistent NEGF-Poisson loop (v1--v2) to acknowledging the semi-analytical nature of the reduced-order Landauer benchmark (v3) and to providing an explicit road map for future self-consistent calculations (v4). This evolution does not invalidate the screening physics; rather, it refines the scope and applicability of the model.

\section{Device Architecture}
\label{sec:architecture}

The vertical MR-DWELL stack consists of a parallel array of sub-micron emitters (overall area $2\times 2~\mathrm{cm}^2$). Each emitter comprises (from bottom to top): an $n^+$-GaAs collector, an $\mathrm{Al}_{0.3}\mathrm{Ga}_{0.7}\mathrm{As}$ barrier ($3~\mathrm{nm}$), an active region with $N_p=10$ DWELL periods, a second $\mathrm{Al}_{0.3}\mathrm{Ga}_{0.7}\mathrm{As}$ barrier ($3~\mathrm{nm}$), an $n^+$-GaAs emitter, an h-BN spacer ($t_d=0.5$--$3~\mathrm{nm}$, $\varepsilon_r\approx 3.9$), and a Graphene/Ti/Au top contact.

Each DWELL period contains a $12~\mathrm{nm}$ $\mathrm{In}_{0.15}\mathrm{Ga}_{0.85}\mathrm{As}$ well with self-assembled InAs quantum dots (nominal base diameter $20~\mathrm{nm}$, height $5~\mathrm{nm}$, density $4\times 10^{10}~\mathrm{cm}^{-2}$), separated by $3~\mathrm{nm}$ barriers. The elementary emitter area is $a_{\mathrm{emit}}\approx 10^{-10}~\mathrm{cm}^2$; the total channel count for a~$2\times 2~\mathrm{cm}^2$ die is $N\sim 4\times 10^{10}$. All channels are parallel; $I_{\mathrm{total}}=N\cdot I_{\mathrm{single}}$.

\begin{figure}[!ht]
\centering
\includegraphics[width=0.49\textwidth]{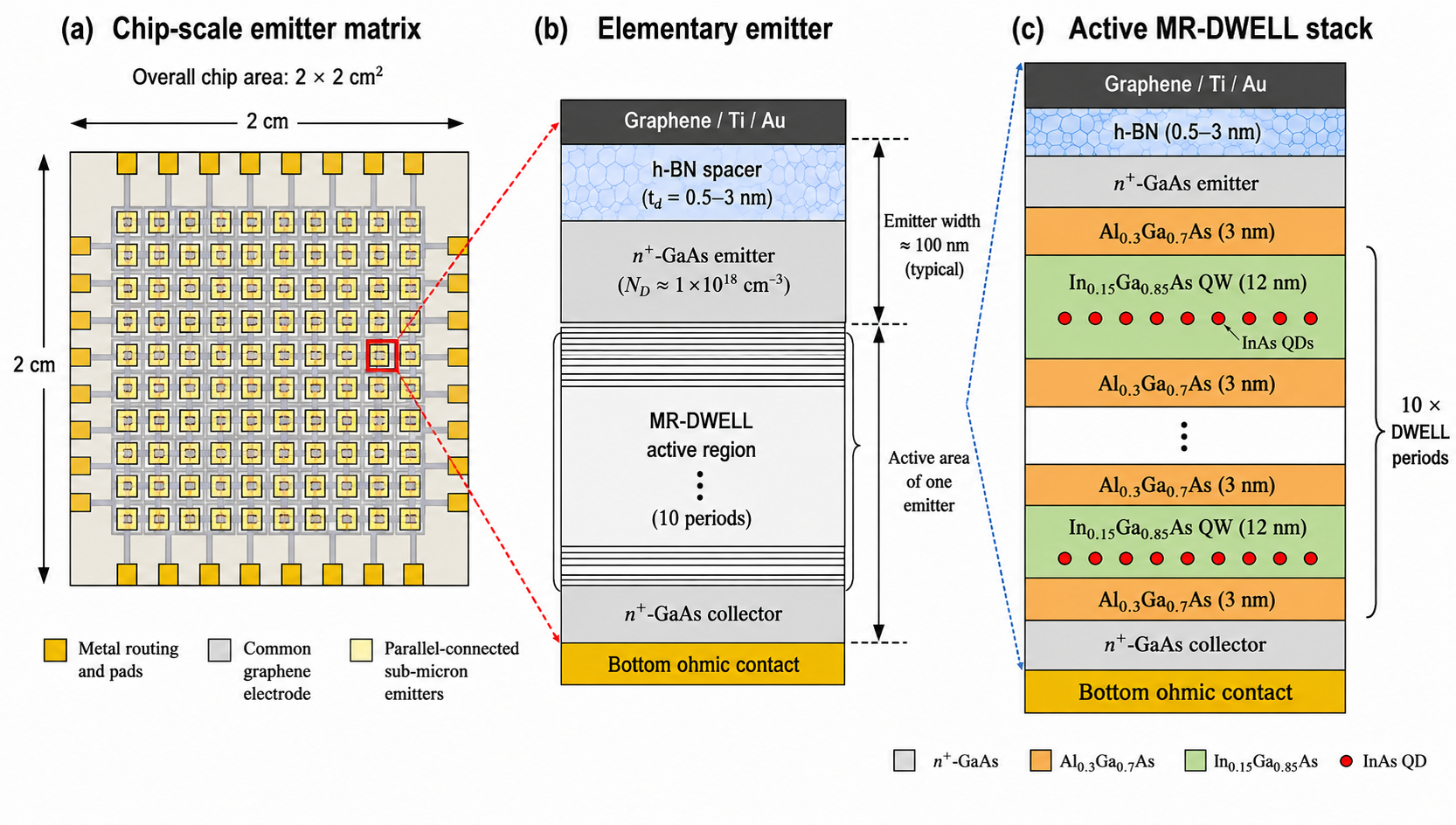}
\caption{\small Schematic of the elementary emitter layer sequence, illustrating the vertical DWELL resonant-tunneling structure with the integrated graphene top contact and h-BN dielectric spacer. (a)~Chip-scale emitter matrix ($2\times 2~\mathrm{cm}^2$). (b)~Elementary emitter cross-section. (c)~Active MR-DWELL stack detail showing the 10-period quantum-dot active region.}
\label{fig:architecture}
\end{figure}

\begin{figure}[!ht]
\centering
\includegraphics[width=0.49\textwidth]{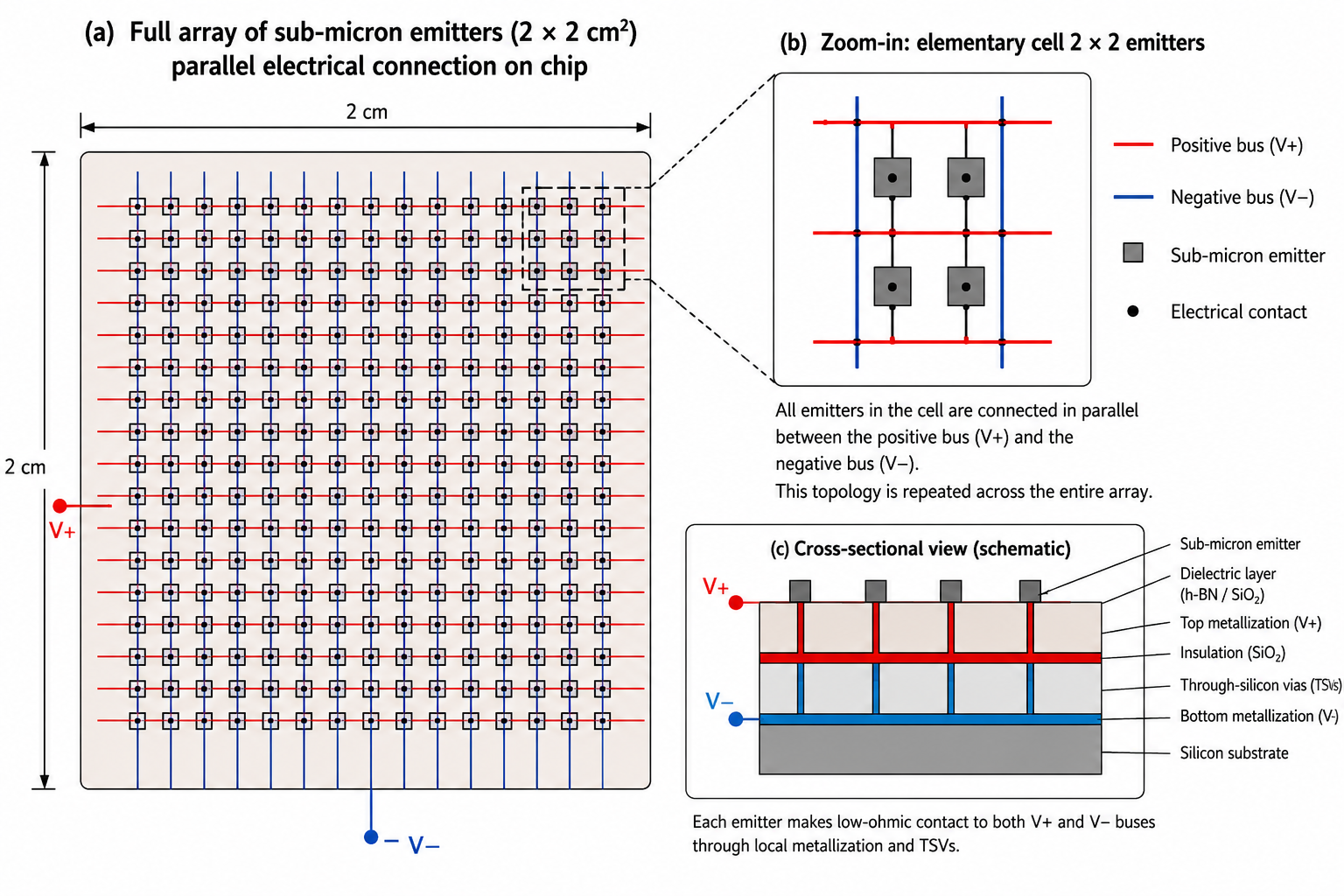}
\caption{\small Conceptual architecture of the proposed MR-DWELL detector array. (a)~Full $2\times 2~\mathrm{cm}^2$ array of parallel-connected sub-micron emitters sharing common positive (V$+$) and negative (V$-$) buses. (b)~Enlarged view of an elementary $2\times 2$ emitter cell illustrating the parallel electrical interconnection. (c)~Schematic cross-sectional view showing the multilayer integration with top and bottom metallisation, dielectric insulation, and through-silicon vias (TSVs).}
\label{fig:array}
\end{figure}

\begin{figure}[!hb]
\centering
\includegraphics[width=0.49\textwidth]{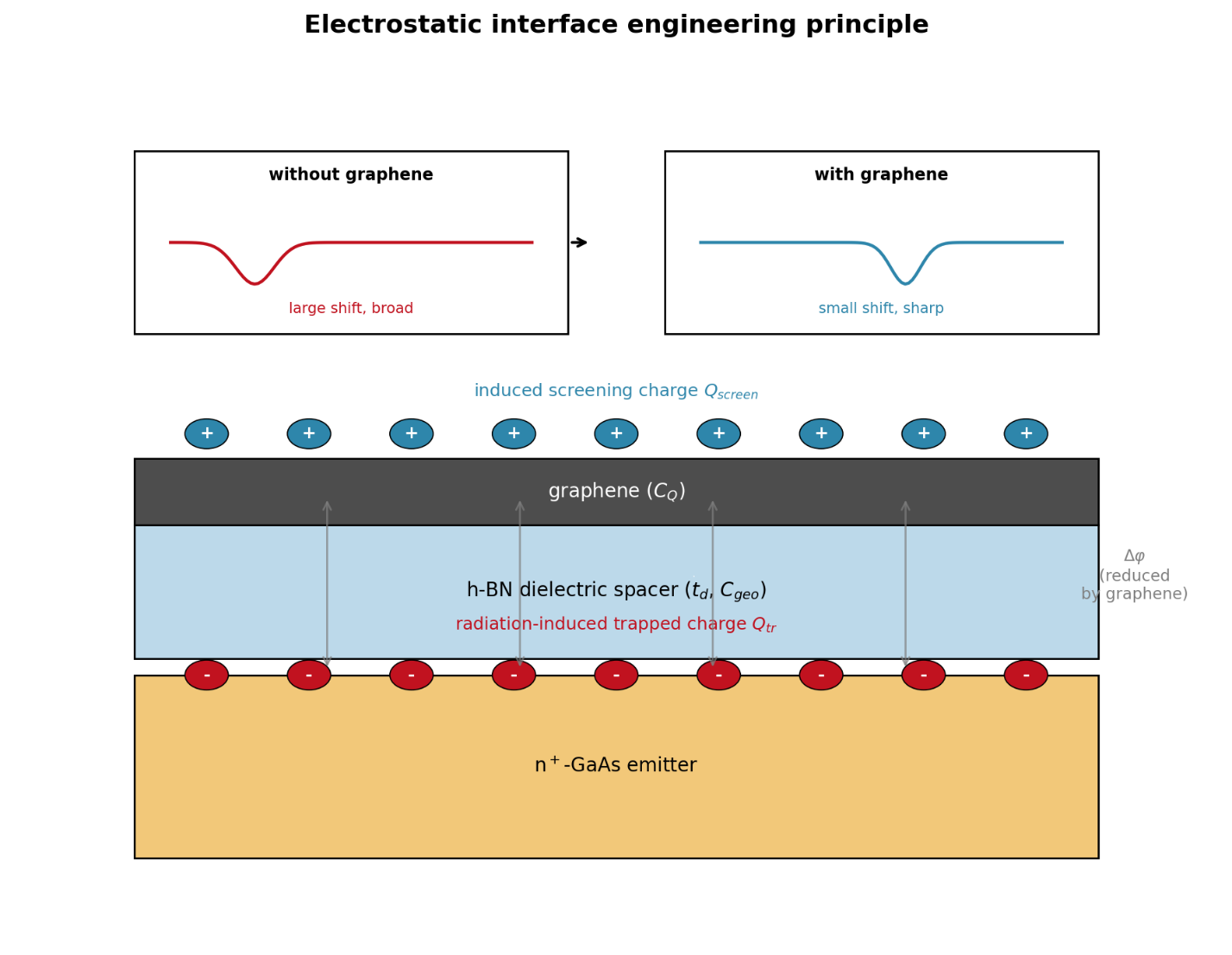}
\caption{\small Electrostatic interface engineering principle demonstrating the screening of radiation-induced trapped charge by the graphene quantum capacitance, stabilising the potential and resonance levels, governed by the screening factor $S=\eta/(1+\eta)$.}
\label{fig:principle}
\end{figure}

The layer sequence is illustrated in Fig.~\ref{fig:architecture}, while the complete array of sub-micron emitters is depicted in Fig.~\ref{fig:array}. Figure~\ref{fig:principle} presents the electrostatic interface engineering principle.

The geometric capacitance of the h-BN spacer is
\begin{equation}
C_{\mathrm{geo}} = \frac{\varepsilon_0 \varepsilon_r a_{\mathrm{emit}}}{t_d},
\label{eq:cgeo}
\end{equation}
and the graphene quantum capacitance (for $E_F\approx 0.3~\mathrm{eV}$) is $C_Q \approx 0.72~\mathrm{fF}$ per emitter. The dimensionless design parameter is
\begin{equation}
\eta = \frac{C_Q}{C_{\mathrm{geo}}}, \qquad S = \frac{\eta}{1+\eta}.
\label{eq:eta}
\end{equation}
The screening factor $S$ represents the fraction of the trapped-charge potential compensated by graphene. For~$\eta \gg 1$, screening is strong ($S\to 1$); for $\eta \ll 1$, it is weak.

\section{Theoretical Model}
\label{sec:model}

\subsection{Phenomenological trapped-charge kinetics}
\label{sec:kinetics}

Radiation-induced trapped charge follows first-order saturation:
\begin{equation}
Q_{\mathrm{tr}}(D) = Q_{\max}\left(1 - e^{-D/D_0}\right), \quad D_0 = 500~\mathrm{Gy}.
\label{eq:trap_kinetics}
\end{equation}
We assume that radiation-generated holes are trapped primarily at the AlGaAs/GaAs interfaces and in the barrier regions, rather than inside the quantum dots themselves, consistent with the RADFET-like trapping model. This assumption avoids the fast recombination expected within the dot active region.

Values are adopted from established RADFET literature~\cite{Sze2006} and serve as conservative estimates;
actual trapping efficiencies may be refined through direct experimental calibration.

The resulting resonance-energy shift and linewidth broadening are modeled as
\begin{align}
E_i(D) &= E_{i0} - \Delta E_{i,\max}\left(1 - e^{-D/D_0}\right), \label{eq:energy_shift}\\
\Gamma_i(D) &= \Gamma_{i0} + \Delta\Gamma_{i,\max}\left(1 - e^{-D/D_0}\right). \label{eq:linewidth}
\end{align}
For the unscreened reference $\Delta E_{1,\max}^{(0)}\approx 70~\mathrm{meV}$; with graphene screening
\begin{equation}
\Delta E_{i,\max}(\eta) = \frac{\Delta E_{i,\max}^{(0)}}{1+\eta}.
\label{eq:screened_shift}
\end{equation}
The linewidth parameters are $\Gamma_{10}=4~\mathrm{meV}$, $\Gamma_{20}=7~\mathrm{meV}$, $\Delta\Gamma_{1,\max}=0.40~\mathrm{meV}$, $\Delta\Gamma_{2,\max}=0.25~\mathrm{meV}$.

This computed energy \(E_0 \approx 0.25\,\mathrm{eV}\) serves as the reference level for the representative resonant state in the NEGF Hamiltonian, relative to which the radiation-induced shift \(\Delta E_{\mathrm{res}}(D)\) is applied to the conduction-band profile. As established for the identical nominal geometry in Ref.~\cite{Dosymova2026d}, the strongly dot-localized ground state lies below the emitter transport edge and is not itself injected into; the accessible sub-barrier resonances that carry the current, and that are shifted by trapped charge, are quantum-well-derived states for which the InAs dots set the local confinement landscape rather than acting as the conducting channel. The representative-level treatment used below should accordingly be understood as describing this quantum-well-derived transport channel.

\subsection{Independent low-dose sensitivity coefficient}
\label{sec:sensitivity}

The low-dose shift coefficient $\alpha_i^{\mathrm{eff}}$ is treated as an independently calibrated descriptor extracted from the sub-mGy transport response. It scales with the screening factor as
\begin{equation}
\alpha_i^{\mathrm{eff}} = \frac{\alpha_i}{1+\eta}, \quad \alpha_1 = 0.14~\mathrm{meV/mGy}.
\label{eq:alpha_eff}
\end{equation}
Importantly, $\alpha_i^{\mathrm{eff}}$ is \emph{not} identified with the mathematical derivative $\Delta E_{i,\max}/D_0$ of the saturation law. The~two quantities characterize different physical limits: $\alpha_i^{\mathrm{eff}}$ governs the observable current sensitivity at $D\ll D_0$, whereas $\Delta E_{i,\max}$ and $D_0$ describe the asymptotic saturation of the electrostatic shift.

\subsection{Landauer transport}
\label{sec:landauer}

The coherent tunneling current is~\cite{Datta1995}:
\begin{equation}
I_{\mathrm{single}}(V) = \frac{2q}{h}\int T(E,D)\bigl[f_L(E) - f_R(E)\bigr]\,dE,
\label{eq:landauer}
\end{equation}
where
\begin{equation}
T(E,D) = \sum_i \frac{\Gamma_{Li}(D)\Gamma_{Ri}(D)}{\bigl(E-E_i(D)\bigr)^2 + \Gamma_i(D)^2},
\label{eq:transmission}
\end{equation}
with $\Gamma_i=(\Gamma_{Li}+\Gamma_{Ri})/2$ within the reduced analytical transport model used here. Two representative literature-calibrated resonances are used
($E_{10}=82~\mathrm{meV}$, $\Gamma_{10}=4~\mathrm{meV}$;
$E_{20}=126~\mathrm{meV}$, $\Gamma_{20}=7~\mathrm{meV}$),
with the resonance-energy scale motivated by experimental studies of resonant tunneling through InAs quantum-dot structures~\cite{Itskevich1996,Narihiro1997}.

\paragraph{Numerical implementation.} 
The one-dimensional transport axis is discretized with a uniform step of 0.5 nm over the 180 nm total modelled structure, comprising the 150 nm active DWELL stack (ten 15-nm periods, each a~12-nm In$_{0.15}$Ga$_{0.85}$As well containing the 5-nm InAs dot segment plus a 3-nm Al$_{0.3}$Ga$_{0.7}$As barrier) together with 15-nm GaAs access regions on each side. Note that the two external 3-nm Al$_{0.3}$Ga$_{0.7}$As barriers of the full device stack are treated as boundary conditions and are not included in the 180 nm transport domain; the total physical stack height is therefore 186 nm. Open boundary conditions are implemented via the recursive Green's function (RGF) method, yielding \(O(N_z)\) scaling. Open-boundary transport is formulated for a one-dimensional open quantum system coupled to particle reservoirs~\cite{Frensley1990}, and is implemented numerically using the recursive Green's function method~\cite{LakeDatta1992}. The~radiation-induced electrostatic perturbation of Eqs.~\eqref{eq:energy_shift}--\eqref{eq:linewidth} is prescribed externally rather than obtained from a converged self-consistent Poisson loop; no potential-convergence criterion is therefore imposed on this step, consistent with the semi-analytical NEGF/Landauer characterization used throughout this work and in Ref.~\cite{Dosymova2026d}. The Hamiltonian is constructed using an effective-mass approximation with material parameters adopted from standard III–V literature~\cite{Sze2006} (in~particular, the specific InAs/InGaAs/GaAs band parameters used for this geometry are consistent with the companion NEGF analysis of Ref.~\cite{Dosymova2026d}).

In the present work, Eq.~\eqref{eq:landauer} provides the physical basis of the transport model, whereas the reported peak and valley currents are evaluated using calibrated reduced transport functions derived from this analytical framework. This reduced approach allows efficient parameter-space sweeps while preserving the physical trends of the resonant tunneling mechanism.

\subsection{Graphene quantum capacitance and response time}
\label{sec:quantum_cap}

Within the lumped-capacitance approximation, the~graphene quantum capacitance is modeled as
\begin{equation}
C_Q = e^2\frac{\partial n}{\partial\mu} = \frac{2e^2 k_B T}{\pi(\hbar v_F)^2}\ln\!\left[2\cosh\!\left(\frac{\mu}{2k_B T}\right)\right],
\label{eq:cq}
\end{equation}
which for $E_F\approx 0.3~\mathrm{eV}$ gives $C_Q$ per area $\approx 7.2~\mu\mathrm{F/cm}^2$~\cite{Xia2009}. For $a_{\mathrm{emit}}=10^{-10}~\mathrm{cm}^2$, $C_Q\approx 0.72~\mathrm{fF}$. In the present analysis, spatial inhomogeneity of the graphene Fermi level and disorder-induced puddles are neglected; $C_Q$ is treated as a uniform, bias-tunable quantity parameterised by $\mu$.

With graphene, $\tau_{\mathrm{int}}^{\mathrm{RC}}\approx 0.28~\mathrm{ps}$ versus $\sim 0.41~\mathrm{ps}$ for metal. The system response $\tau_{\mathrm{system}}\approx \tau_{\mathrm{int}}^{\mathrm{RC}} + \tau_{\mathrm{interconnect}} + \tau_{\mathrm{electronics}}$ is estimated at $\sim 20~\mathrm{ps}$, packaging-dependent.

\subsection{Self-consistent charge--transport framework}
\label{sec:selfconsistent}

For quantitative prediction beyond the compact benchmark, the following chain must be solved:
$$ \dot{D}(z,t) \to G(z,t) \to f_{t,j}(z,t) \to $$
\begin{equation}\to \rho(z,t) \to \phi(z,t) \to E_i(t) \to I(V,t).
\label{eq:chain}
\end{equation}
The potential follows the Poisson equation with heterointerface boundary conditions:
\begin{equation}
\frac{d}{dz}\!\left[\varepsilon(z)\frac{d\phi}{dz}\right] = -\rho(z,t),
\label{eq:poisson}
\end{equation}
\begin{equation}
\rho = q\!\left[p - n + N_D^+ - N_A^- + \sum_j s_j N_{t,j} f_{t,j}\right].
\label{eq:charge_density}
\end{equation}
The Hamiltonian $H[\phi]$ yields the retarded Green's function
\begin{equation}
G^r(E) = \bigl[EI - H[\phi] - \Sigma_L^r - \Sigma_R^r - \Sigma_{\mathrm{scatt}}^r\bigr]^{-1},
\label{eq:green}
\end{equation}
and the state-specific shift is
\begin{equation}
\Delta E_i(t) = -q\int |\psi_i(z)|^2\bigl[\phi(z,t) - \phi(z,0)\bigr]\,dz.
\label{eq:state_shift}
\end{equation}
The device-scale current requires ensemble averaging over the QD energy distribution $P_i(E_0)$ because the inhomogeneous broadening ($\sigma_E\approx 10~\mathrm{meV}$) exceeds the intrinsic linewidth:
\begin{equation}
I_{\mathrm{ens}}(V,t) = \sum_{i=1}^{N_p}\int I_i(V,t;E_0)P_i(E_0)\,dE_0.
\label{eq:ensemble}
\end{equation}

\subsection{Second-derivative spectroscopy and detection limit}
\label{sec:spectroscopy}

The readout signal
\begin{equation}
\frac{d^2I}{dV^2} = \frac{d}{dV}\!\left(\frac{dI}{dV}\right)
\label{eq:d2idv2}
\end{equation}
tracks resonance evolution via lock-in techniques~\cite{Britnell2012,Khanin2018}. Inelastic electron tunneling spectroscopy has been experimentally demonstrated in graphene/hBN heterostructures, where second-harmonic detection of $d^2I/dV^2$ reveals vibrational modes~\cite{Jung2015}. The detection limit is readout-conditional:
\begin{equation}
D_{\min} = \frac{k_\sigma\, i_{n,\mathrm{tot}}}{|S_I|},
\label{eq:dmin}
\end{equation}
\begin{equation}
i_{n,\mathrm{tot}}^2 = i_{n,\mathrm{shot}}^2 + i_{n,\mathrm{Johnson}}^2 + i_{n,\mathrm{GR}}^2 + i_{n,1/f}^2 + i_{n,\mathrm{RTN}}^2 + i_{n,\mathrm{amp}}^2,
\label{eq:noise}
\end{equation}
where $S_I = dI/dD|_{D\to 0}$ is the dose sensitivity. Any claim of a microgray-level limit must specify bias, temperature, bandwidth, and all noise contributions. While the analytical model predicts sub-mGy sensitivity, the practical detection limit is ultimately dominated by the amplifier and $1/f$ noise contributions, especially for low-bandwidth, high-precision applications.

\subsection{Model assumptions and limitations}
The present semi-analytical framework relies on several simplifying assumptions. First, the trapping kinetics are described by a phenomenological model (Eq.~\ref{eq:trap_kinetics}) adapted from RADFETs; the actual capture efficiency in DWELL structures may differ due to spatially localized states. Second, the model assumes coherent quantum transport and neglects inelastic scattering (electron-phonon interaction), which becomes relevant at room temperature and would lead to additional linewidth broadening. Third, trap-assisted tunneling and charge re-emission are not explicitly included. Despite these idealizations, the model captures the dominant electrostatic modulation pathway: ionization \(\rightarrow Q_{\mathrm{tr}} \rightarrow \Delta V \rightarrow \Delta E_{\mathrm{res}} \rightarrow I\), and provides a~physically consistent upper bound for the low-dose linear response.

Two further points require explicit cross-reference to the companion revision of the base transport model for the identical nominal geometry~\cite{Dosymova2026d}. (i) The resonance parameters used here, $E_{10}=82$~meV and $E_{20}=126$~meV, are adopted from the literature on comparable DWELL structures~\cite{Itskevich1996,Narihiro1997} rather than computed self-consistently for the present 10-period, 3-nm-barrier stack; a direct NEGF calculation for this nominal geometry instead places the two hybridized resonance groups near $\sim\!44$~meV and $\sim\!173$~meV, with individual widths that hybridize into minibands of order $18$ and $53$~meV. The two-level Breit-Wigner form of Eq.~\eqref{eq:transmission} should therefore be read as a~reduced, literature-calibrated benchmark for the screening physics ($\eta$, $S$, $\alpha_i^{\mathrm{eff}}$) rather than as a self-consistently derived spectrum for this specific geometry; reconciling the two parameterizations is left for future work. (ii) At the bias required to drive appreciable current through this stack, the same geometry is shown in Ref.~\cite{Dosymova2026d} to enter a field-detuned Wannier--Stark regime in which $k_BT$ and the inter-period potential drop both exceed the zero-bias miniband width, so the room-temperature current is a~thermally averaged quantity rather than a sequence of resolved single-level peaks. The present compact model, which tracks two fixed resonance energies rather than a~bias- and temperature-dependent ladder, is accordingly best interpreted as characterizing the electrostatic screening trend ($S$, $\alpha_i^{\mathrm{eff}}$, PVR versus $\eta$) rather than as a quantitative prediction of absolute peak positions or currents. We note that localized defect states in hBN barriers are a universal phenomenon that can significantly affect resonant tunneling transport. Recent experiments on van der Waals heterostructures have demonstrated that inelastic tunneling through adjacent localized states in hBN can dominate the current-voltage characteristics~\cite{Vdovin2025}. This underscores the importance of understanding the role of such states in the electrostatic environment of the graphene-assisted MR-DWELL architecture, although the present model treats the hBN spacer as an ideal dielectric.

\section{Results and Discussion}
\label{sec:results}

We emphasize that all quantitative performance metrics (sensitivity, minimum detectable dose, and operational range) reported in this work are theoretical predictions derived from the analytical and semi-analytical models. These predictions serve to establish the physical feasibility of the mechanism and require experimental validation. Table~\ref{tab:comparison} provides a conceptual comparison with representative radiation-detection technologies.

\begin{table*}[!ht]
\centering
\caption{Comparison of radiation detection technologies with the proposed MR-DWELL. Values for other devices are literature-based; MR-DWELL metrics are theoretical predictions.}
\label{tab:comparison}
\begin{tabular}{@{}l|c|c|c|c@{}}
\toprule
Device & Principle & Sensitivity & Dose range & \(D_{\min}\) \\
\midrule
RADFET & Threshold shift & 1–10 mV/Gy & 0.1 Gy–10 kGy & \(\sim\)0.1 Gy \\
Si diode & Ionization current & \(\sim\)0.1 $\mu$A/Gy & 0.1 mGy–10 Gy & \(\sim\)0.1 mGy \\
Ionization chamber & Collected charge & \(\sim\)10 fC/Gy & 1 $\mu$Gy–1 Gy & \(\sim\)1 $\mu$Gy \\
\hline
\makecell[l]{{\bf MR-DWELL}\\ (this work)} & Resonance shift & $\approx0.10~\mu$A/mGy (scaling estimate) & 
Readout-dependent; see Eq.~\eqref{eq:dmin} & see~Eq.~\eqref{eq:dmin} \\
\bottomrule
\end{tabular}
\end{table*}

\subsection{Influence of dielectric thickness on screening}
\label{sec:screening}

Table~\ref{tab:screening} summarises capacitance-matching parameters for a single elementary channel.

\begin{table}[!ht]
\centering
\caption{Capacitance-matching parameters for a~single elementary channel ($a_{\mathrm{emit}}=10^{-10}~\mathrm{cm}^2$, $\varepsilon_r=3.9$). The~low-dose shift coefficients $\alpha_i^{\mathrm{eff}}$ are given in meV/mGy; the characteristic dose $D_0=500~\mathrm{Gy}$ governs the approach to the asymptotic shift.}
\label{tab:screening}
\begin{tabular}{ccccc}
\toprule
$t_d$ (nm) & $C_{\mathrm{geo}}$ (fF) & $\eta$ & $S$ & $\alpha^{\mathrm{eff}}$ (meV/mGy) \\
\midrule
0.5 & 0.691 & 1.04 & 0.51 & 0.069 \\
1.0 & 0.345 & 2.09 & 0.68 & 0.045 \\
2.0 & 0.173 & 4.16 & 0.81 & 0.027 \\
3.0 & 0.115 & 6.26 & 0.86 & 0.019 \\
\bottomrule
\end{tabular}
\end{table}

The results in Table~\ref{tab:screening} illustrate the design trade-off between high sensitivity and radiation tolerance: thinner h-BN spacers enhance sensitivity (larger $\alpha^{\mathrm{eff}}$) at the cost of stronger radiation-induced degradation, whereas thicker spacers improve stability at reduced sensitivity.

\subsection{Design regimes}
\label{sec:design}

\begin{figure}[htbp]
\centering
\includegraphics[width=0.49\textwidth]{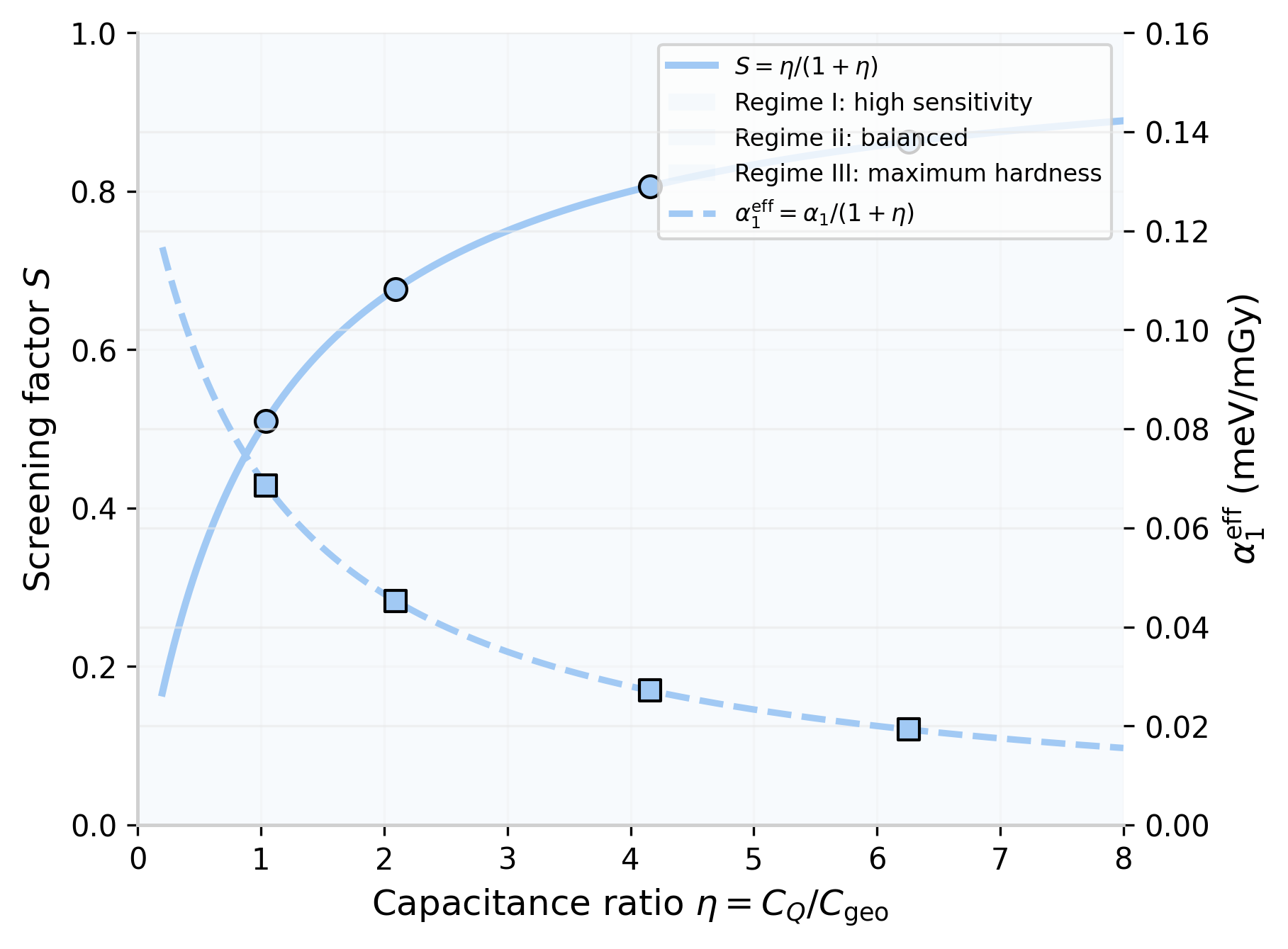}
\caption{\small Design metrics as a function of the capacitance ratio $\eta$. The solid line shows the screening factor $S=\eta/(1+\eta)$ (left axis); the dashed line shows the effective low-dose shift coefficient $\alpha_1^{\mathrm{eff}}=\alpha_1/(1+\eta)$ (right axis). Shaded regions indicate the three design regimes: I (high sensitivity), II (balanced), and III (maximum hardness). Discrete points correspond to the tabulated spacer thicknesses of Table~\ref{tab:screening}.}
\label{fig:design_metrics}
\end{figure}

Figure~\ref{fig:design_metrics} shows the screening factor $S(\eta)$ and the low-dose coefficient $\alpha_1^{\mathrm{eff}}(\eta)$. Three design regimes emerge:
\begin{itemize}
    \item {Regime I} (High sensitivity, $\eta<1.5$): weak screening, $\alpha_1^{\mathrm{eff}}\gtrsim 0.06~\mathrm{meV/mGy}$, pronounced current degradation at $1~\mathrm{MGy}$.
    \item {Regime II} (Balanced, $1.5<\eta<4$): intermediate screening, $\alpha_1^{\mathrm{eff}}\sim 0.03$--$0.05~\mathrm{meV/mGy}$, intrinsic response $\tau_{\mathrm{int}}\approx 0.28~\mathrm{ps}$. This is the recommended window for FLASH dosimetry.
    \item {Regime III} (Maximum hardness, $\eta>4$): strong screening, $\alpha_1^{\mathrm{eff}}\lesssim 0.03~\mathrm{meV/mGy}$, $<10\%$ current drop at $1~\mathrm{MGy}$, reduced sensitivity.
\end{itemize}

\subsection{Resonance shift versus dose}
\label{sec:shift}

\begin{figure}[htbp]
\centering
\includegraphics[width=0.49\textwidth]{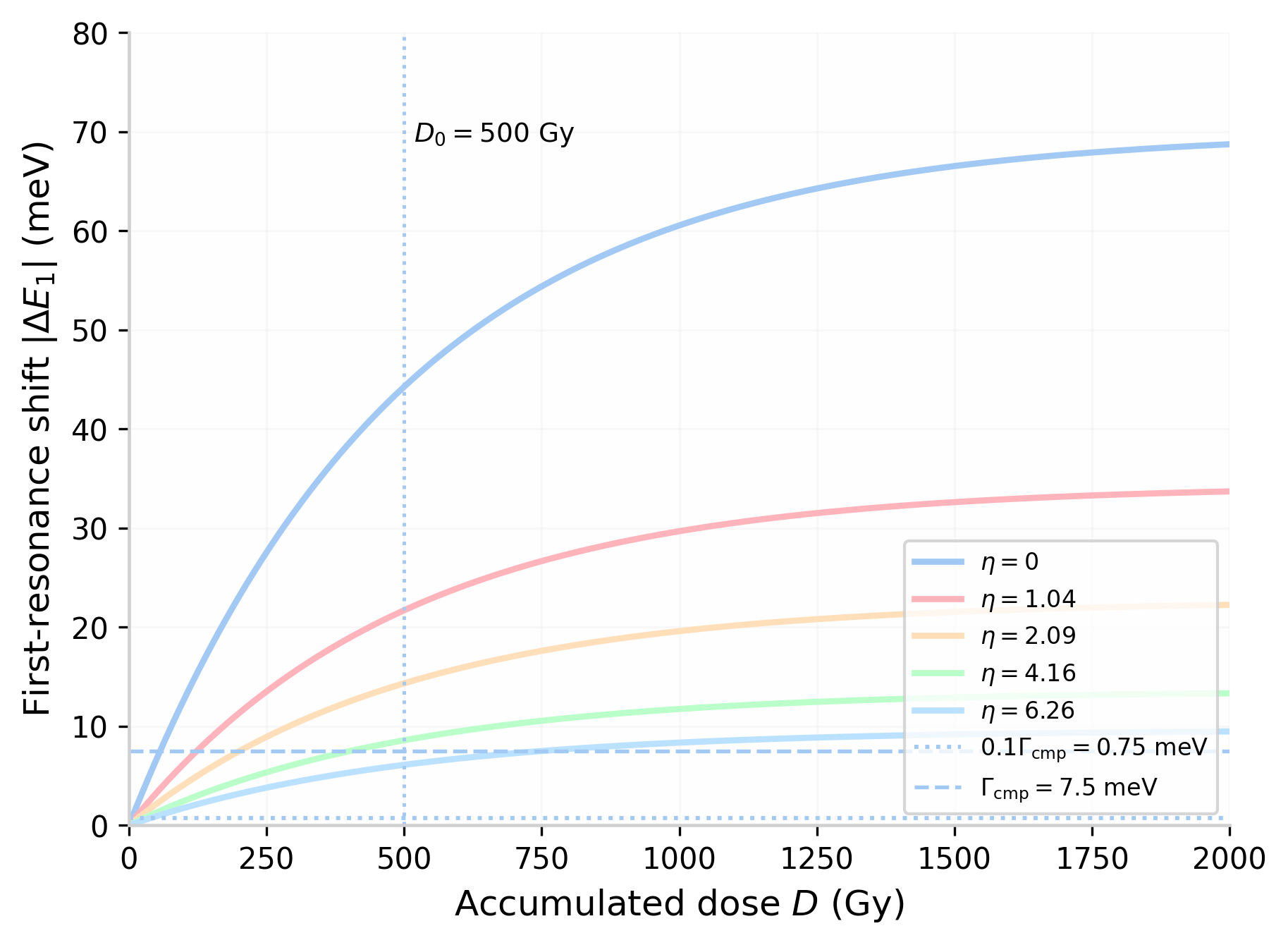}
\caption{\small First-resonance shift $|\Delta E_1|$ versus accumulated dose $D$ for several screening parameters $\eta$. The dotted horizontal line marks the $10\%$-nonlinearity threshold $0.1\Gamma_{\mathrm{cmp}}=0.75~\mathrm{meV}$; the dashed line marks $\Gamma_{\mathrm{cmp}}=7.5~\mathrm{meV}$. The vertical dotted line indicates $D_0=500~\mathrm{Gy}$.}
\label{fig:resonance_shift}
\end{figure}

Figure~\ref{fig:resonance_shift} displays the dose-dependent shift $|\Delta E_1|$ for several $\eta$. At $\eta=2.09$ the shift saturates near $23~\mathrm{meV}$, well below the unscreened $70~\mathrm{meV}$. The horizontal lines mark the compact-model benchmark linewidth $\Gamma=7.5~\mathrm{meV}$ and its $10\%$ fraction. This value is a phenomenological benchmark chosen for the small-perturbation analysis; it~is close to the asymptotic \(\Gamma_2\) value of \(7.25~\mathrm{meV}\) and represents a typical resonance width in the compact model.

Using the dose-dependent energy shift defined by Eq.~\eqref{eq:energy_shift}, $\Delta E_{\max}(D)=22.65\left[1-\exp(-D/D_0)\right]~\mathrm{meV}$ with $D_0=500~\mathrm{Gy}$, the doses corresponding to the reference broadening levels $0.1\Gamma$ and $\Gamma$, where $\Gamma=7.5~\mathrm{meV}$, are obtained directly from the same dose-dependent relation used in Fig.~\ref{fig:resonance_shift}. This gives approximately $D_{0.1\Gamma}\approx 16.8~\mathrm{Gy}$ and $D_{\Gamma}\approx 201~\mathrm{Gy}$. These values therefore represent the dose scale at which the dose-dependent resonance shift reaches $10\%$ and $100\%$ of the reference linewidth, respectively. The resulting doses are appropriate for assessing the validity of the small-shift criterion used in Fig.~\ref{fig:resonance_shift}. A separate low-dose description introduced through the effective coefficient $\alpha^{\mathrm{eff}}$ gives a substantially different numerical scale; it should be regarded as an independent local sensitivity descriptor rather than as an alternative inversion of Eq.~\eqref{eq:energy_shift}. In particular, the $\alpha^{\mathrm{eff}}$-based estimates are not the intersection doses of the curves shown in Fig.~\ref{fig:resonance_shift} and should not be used interchangeably with the values above. The distinction matters because the compact low-dose benchmark and the transport-based dose law describe different parameterizations of the response; the present formulation therefore does not claim to eliminate the numerical scale difference between these descriptors, but keeps them explicitly separated to avoid mixing two different dose scales. Thus, extrapolation beyond the saturation regime without self-consistent electrostatics is invalid.

\subsection{Linewidth evolution}
\label{sec:linewidth}

\begin{figure}[!ht]
\centering
\includegraphics[width=0.49\textwidth]{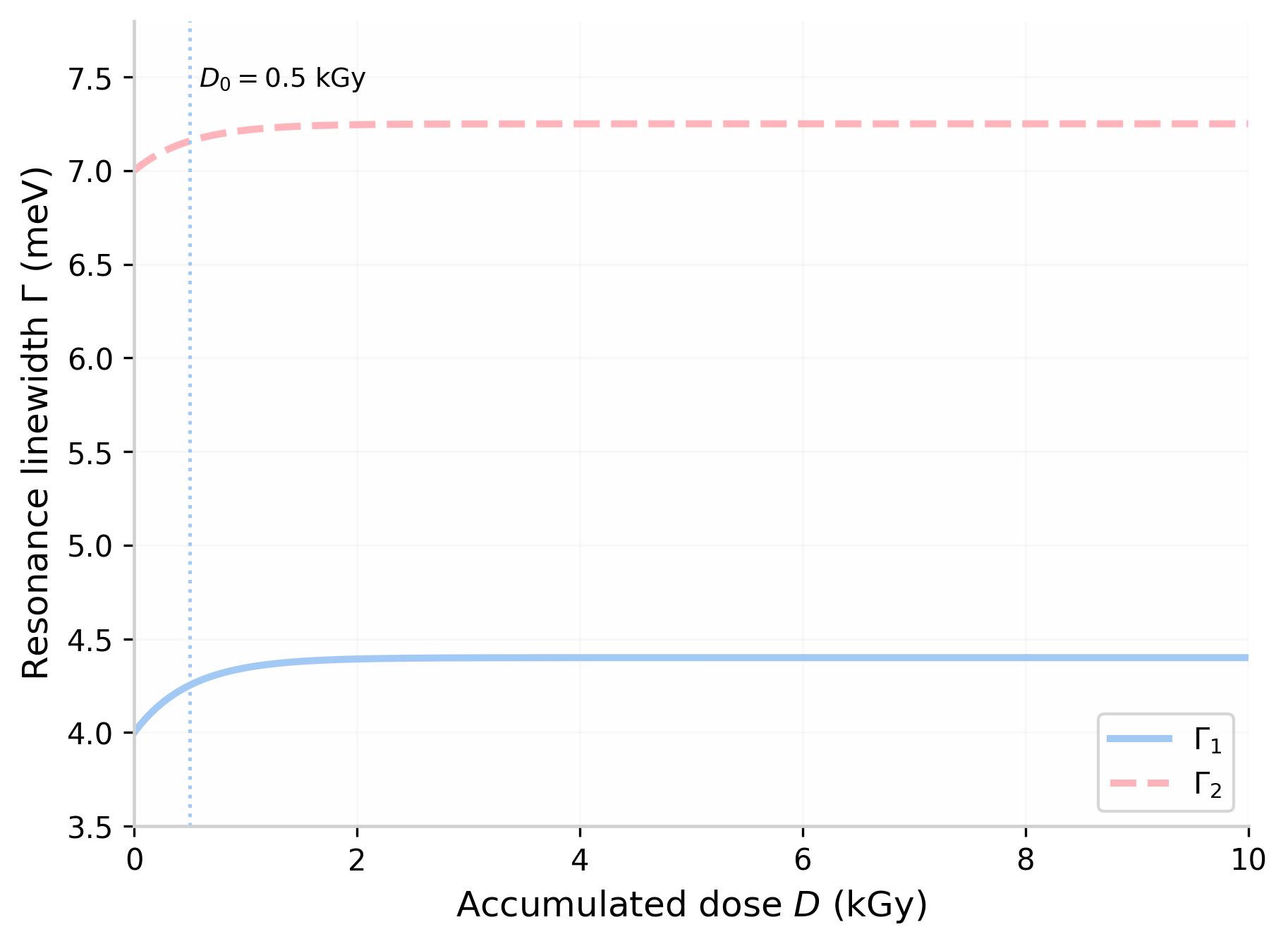}
\caption{\small Resonance linewidths $\Gamma_1$ (solid) and $\Gamma_2$ (dashed) as functions of accumulated dose, calculated from Eq.~\eqref{eq:linewidth}. The~vertical dotted line marks $D_0=0.5~\mathrm{kGy}$.}
\label{fig:linewidth}
\end{figure}

Figure~\ref{fig:linewidth} presents the radiation-induced broadening for the two resonances in the graphene-assisted device ($\eta=2.09$). Both $\Gamma_1$ and $\Gamma_2$ follow the unified exponential saturation model with $D_0=500~\mathrm{Gy}$, approaching asymptotic values of $4.40~\mathrm{meV}$ and $7.25~\mathrm{meV}$, respectively. The~low-dose broadening coefficients are $\beta_1=0.8~\mathrm{meV/kGy}$ and $\beta_2=0.5~\mathrm{meV/kGy}$.

\subsection{Current degradation and spectroscopic response}
\label{sec:degradation}

Table~\ref{tab:degradation} summarizes the analytical performance metrics for representative values of the electrostatic screening parameter $\eta$. Since the screening factor $S=\eta/(1+\eta)$ depends only on $\eta$, the latter serves as a compact design parameter for interface optimization.

\begin{table}[!ht]
\centering
\caption{Representative theoretical performance metrics obtained from the electrostatic-screening model and the reduced Landauer transport model.}
\label{tab:degradation}
\begin{tabular}{ccccc}
\toprule
$\eta$ & $S$ & $\alpha^{\mathrm{eff}}$ (meV/mGy) & Current drop (\%) & 
\(\Delta E_{\mathrm{peak}}\) (meV) \\
\midrule
1.04 & 0.51 & 0.069 & 33 & 42 \\
2.09 & 0.68 & 0.045 & 17 & 25 \\
4.16 & 0.81 & 0.027 & 10 & 15 \\
6.26 & 0.86 & 0.019 & 6  & 8  \\
\bottomrule
\end{tabular}

\medskip
\parbox{\linewidth}{%
\footnotesize
\(S\) and \(\alpha_1^{\mathrm{eff}}\) are obtained analytically from the electrostatic-screening model. The current degradation and \(\Delta E_{\mathrm{peak}}\) are discrete outputs of the reduced Landauer transport model evaluated at \(D=1\) MGy and the listed \(\eta\) values. \(\Delta E_{\mathrm{peak}}\)  
denotes the shift of the spectroscopic extremum obtained from \(d^2I/dV^2\) and is not identical to the bare electrostatic resonance shift \(\Delta E_{1,\max}=70/(1+\eta)\).%
}
\end{table}

The sequence of current peaks and subsequent drops in the I-V characteristics corresponds to the negative differential conductance (NDC) typical of resonant-tunneling structures. In the present work, we focus primarily on the dose-dependent shift of the resonance, while the NDC feature is preserved in the transport response.

\begin{figure}[!ht]
\centering
\includegraphics[width=0.48\textwidth]{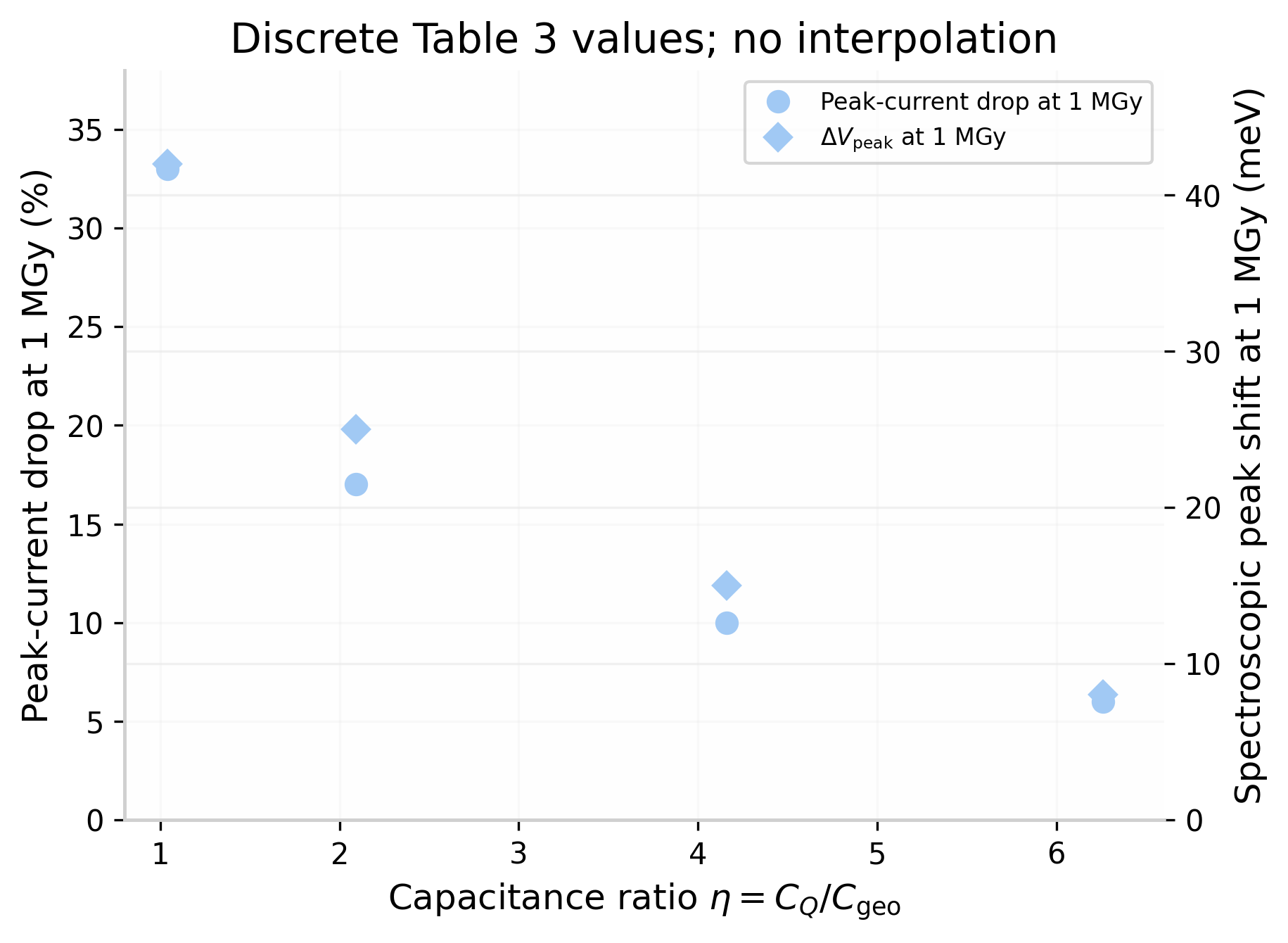}
\caption{\small Discrete model outputs at $1~\mathrm{MGy}$: peak-current drop (circles, left axis) and spectroscopic peak shift $\Delta E_{\mathrm{peak}}$
(diamonds, right axis) versus the screening parameter $\eta$. No~interpolation is applied.}
\label{fig:table3}
\end{figure}

Figure~\ref{fig:table3} gives the model-predicted peak-current degradation and $d^2I/dV^2$ peak shift at $1~\mathrm{MGy}$ as a function of $\eta$. These are discrete model outputs; no continuous interpolation is imposed because the underlying transport calculation was performed only at the tabulated $\eta$ points. The trend confirms that increasing $\eta$ monotonically suppresses both current loss and spectroscopic shift.

\subsection{Operating map}
\label{sec:map}

\begin{figure}[htbp]
\centering
\includegraphics[width=0.49\textwidth]{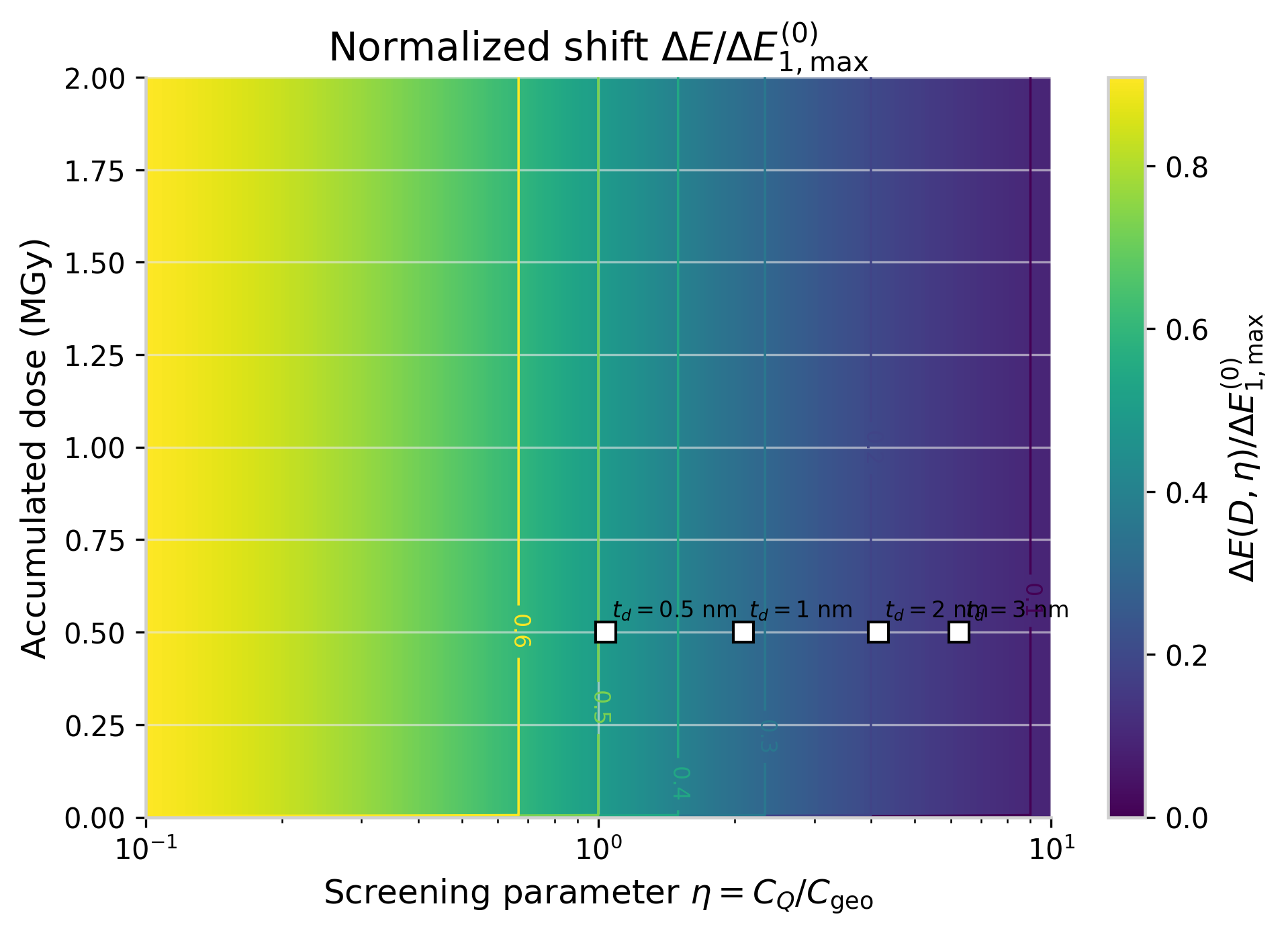}
\caption{\small Two-dimensional operating map: normalized resonance shift $\Delta E(D,\eta)/\Delta E_{1,\max}^{(0)}$ as a function of screening parameter $\eta$ and accumulated dose $D$. Contour lines are drawn at $0.1$, $0.2$, $0.3$, $0.4$, $0.5$, and $0.6$. White squares mark the spacer thicknesses of Table~\ref{tab:screening}.}
\label{fig:operating_map}
\end{figure}

Figure~\ref{fig:operating_map} shows the two-dimensional design map of the normalized resonance shift
\begin{equation}
\frac{\Delta E(D,\eta)}{\Delta E_{1,\max}^{(0)}} = \frac{1}{1+\eta}\left(1 - e^{-D/D_0}\right).
\label{eq:normalized_shift}
\end{equation}
The contours allow direct selection of $\eta$ for a prescribed dose budget. For example, to keep the normalized shift below $0.3$ at $0.5~\mathrm{MGy}$ requires $\eta\gtrsim 2$.

\subsection{$d^2I/dV^2$ spectroscopic analysis}
\label{sec:d2idv2}

To evaluate experimentally accessible signatures of radiation-induced resonance evolution, the second derivative of the current--voltage characteristic ($d^2I/dV^2$) was calculated using the same transport model. Unlike the transmission probability, the $d^2I/dV^2$ signal can be measured directly using standard lock-in techniques and therefore provides a practical approach for tracking resonance positions. Figure~\ref{fig:table3} reports the resulting spectroscopic peak shift 
$\Delta E_{\mathrm{peak}}$ at $1~\mathrm{MGy}$ as a function of $\eta$, showing that the shift decreases monotonically with increasing screening; a dose-resolved plot of the $d^2I/dV^2$ spectra themselves is not included in the present version. These results support the use of differential-conductance spectroscopy in the proposed model.

In a practical implementation, $d^2I/dV^2$ can be extracted using a two-stage lock-in scheme with a small AC modulation superimposed on the DC bias across the MR-DWELL stack. The modulation frequency is chosen above the $1/f$ noise corner of the front-end but below any parasitic resonances of the interconnects, ensuring that the intrinsic response time of the device does not impose additional bandwidth constraints.

\subsection{Peak-to-valley ratio and temporal response}
\label{sec:pvr}

The peak-to-valley ratio is obtained from the transport model:
\begin{equation}
\mathrm{PVR}(D,\eta) = \frac{I_{\mathrm{peak}}(D,\eta)}{I_{\mathrm{valley}}(D,\eta)},
\label{eq:pvr}
\end{equation}
where the peak and valley currents are computed from the reduced transport functions. At $\eta=2.09$, PVR remains $\sim 10$ at $1~\mathrm{MGy}$.

\subsection{Robustness}
\label{sec:robustness}

To evaluate the stability of the proposed design framework against structural and electrostatic variations, a~comprehensive set of 108 parameter-sweep simulations covering 3 barrier thicknesses, 12 interface trap densities, and 3 quantum capacitance values was performed. Over the full swept range, the effective shift coefficient $\alpha^{\mathrm{eff}}$ varies by $-10.6\%/+9.5\%$ for $t_b$ ($2.4$--$3.6~\mathrm{nm}$), $-16.7\%/+20.0\%$ for $D_{\mathrm{it}}$ ($10^{10}$--$10^{12}~\mathrm{cm}^{-2}\mathrm{eV}^{-1}$), and $-16.9\%/+25.4\%$ for $C_Q$ ($\pm 30\%$); the peak current $J_p$ varies by $-55\%/+122\%$, $-68\%/+216\%$, and $-6.6\%/+4.4\%$ respectively; while the PVR varies by $-21\%/+27\%$, $-37\%/+58\%$, and $-8.2\%/+5.5\%$, respectively.

\subsection{Performance comparison: metal versus graphene interface}
\label{sec:comparison}

Table~\ref{tab:metal_comparison} compares the proposed graphene-assisted interface ($\eta=2.09$) with the reference unscreened architecture~\cite{Dosymova2026a} and an ideal metal-on-h-BN contact (extrapolated using the same capacitive model).

\begin{table}[!ht]
\centering
\caption{Comparison of the proposed graphene-assisted interface with the unscreened architecture~\cite{Dosymova2026a} and an ideal metal contact.}
\label{tab:metal_comparison}
\footnotesize
\setlength{\tabcolsep}{1pt}
\renewcommand{\arraystretch}{1.05}
\begin{tabular}{@{}l|c|c|c@{}}
\toprule
Parameter &
Unscreened~\cite{Dosymova2026a} &
Graphene &
\makecell[l]{Ideal\\ metal\\ (extrapolated)} \\
\midrule
$\alpha^{\mathrm{eff}}$ (meV/mGy) &
0.14 &
0.045 &
0 \\
\hline
\makecell[l]{Shift\\ at 1 MGy (meV)} &
\makecell[l]{$\sim70$\\ (degraded)} &
\makecell[l]{$\sim23$\\ (preserved)} &
$\sim0$ \\
\hline
PVR at 1 MGy &
$\sim1.5$ &
$\sim10$ &
$\sim12$ \\
\hline
$\tau_{\mathrm{int}}$ (ps) &
--- &
$\sim0.28$ &
$\sim0.41$ \\
\bottomrule
\end{tabular}
\end{table}

\begin{table*}[!ht]
\centering
\caption{Comparison of representative radiation-detector platforms with the proposed graphene-assisted MR-DWELL dosimeter. The comparison is intended for conceptual positioning rather than direct metrological equivalence because the cited studies use different detector materials, geometries, irradiation conditions, and readout metrics. All performance values for the proposed MR-DWELL are theoretical predictions.}
\label{tab:comparison_platforms}
\footnotesize
\setlength{\tabcolsep}{3pt}
\renewcommand{\arraystretch}{1.1}

\resizebox{\textwidth}{!}{%
\begin{tabular}{@{}lcccc@{}}
\toprule
\textbf{Parameter / FoM} &
\parbox{2.5cm}{\centering\textbf{Graphene/CsPbBr$_3$}} &
\parbox{2.5cm}{\centering\textbf{FAPbBr$_3$ single crystal}} &
\parbox{2.5cm}{\centering\textbf{InAs/GaAs QD structures}} &
\parbox{2.8cm}{\centering\textbf{Proposed graphene-assisted MR-DWELL}} \\
\midrule
\textbf{Representative literature} &
Liu et al.~\cite{Liu2026} &
Jayarathne et al.~\cite{Jayarathne2026} &
Rahaman and Ghosh~\cite{Rahaman2022} &
This work \\
\textbf{Primary design lever} &
Graphene-assisted carrier transport &
Defect / strain engineering &
QD size and bandstructure &
Capacitance ratio $\eta=C_Q/C_{\mathrm{geo}}$ \\
\textbf{Detection / transduction} &
Photoconductive X-ray detection &
Charge generation and collection &
Photo-induced carrier dynamics &
Electrostatic modulation of resonant tunneling \\
\textbf{Target application} &
Low-dose X-ray detection &
Radiation detection / spectroscopy &
IR / NIR photodetection &
FLASH dosimetry and MGy-class operation \\
\textbf{Representative metric} &
X-ray sensitivity / detection limit &
Energy resolution / detection performance &
Responsivity / detectivity &
$S_I\approx0.10~\mu\mathrm{A/mGy}$ (scaling estimate) \\
\textbf{Intrinsic response} &
Carrier transport and recombination &
Carrier generation and extraction &
Carrier dynamics &
Radiation-induced resonant-state shift and tunneling-current modulation \\
\textbf{Speed} &
Transport / integration limited &
Charge-transport limited &
Carrier-dynamics dependent &
$\tau_{\mathrm{int}}\approx0.28$ ps; $\tau_{\mathrm{sys}}\approx20$ ps (predicted) \\
\textbf{Radiation-hardness evidence} &
Experimental X-ray detector &
Experimental radiation-detector study &
QD irradiation studies &
Theoretical prediction: PVR $\sim10$ and $>80\%$ peak-current retention at 1 MGy \\
\textbf{Readout} &
Conventional detector readout &
Conventional detector readout &
Optical wavelength selectivity &
$d^2I/dV^2$ spectroscopy \\
\textbf{Radiation-hardness basis} &
Graphene-assisted transport &
Reduced defects and strain engineering &
Quantum confinement / localized states &
Electrostatic screening of radiation-induced trapped charge \\
\textbf{What is demonstrated} &
Experimental X-ray detection &
Experimental radiation detection &
Experimental + modelling &
Theoretical concept; no experimental validation \\
\bottomrule
\end{tabular}%
} 
\end{table*}

\emph{Note:} The current sensitivity listed for the graphene-assisted architecture in Tables~\ref{tab:comparison} and \ref{tab:comparison_platforms} is a first-order scaling estimate rather than an independently recalculated transport result; it is therefore included only for indicative comparison. (The value $0.10~\mu$A/mGy is obtained by applying the electrostatic sensitivity reduction $0.045/0.14$ to the previously reported $0.3~\mu$A/mGy; this scaling should not be interpreted as an independent calculation.)

\begin{table*}[!ht]
\centering
\caption{Qualitative comparison of radiation hardness indicators for the graphene-assisted MR-DWELL dosimeter and SiC detectors.}
\label{tab:sic}
\begin{tabular}{lcc}
\toprule
Characteristic & SiC detector & Proposed MR-DWELL \\
\midrule
Radiation tolerance & Demonstrated to 1 MGy under 160-keV X-rays & Predicted stable up to $1~\mathrm{MGy}$ \\
\hline
Dominant degradation & Bulk radiation damage & Electrostatic perturbation of resonant states \\
\hline
Electrostatic stability & High & Enhanced by graphene screening \\
\hline
System response time & --- & $\sim 0.28~\mathrm{ps}$ intrinsic; $\sim 20~\mathrm{ps}$ system-level (predicted) \\
\hline
Energy selectivity & None & Intrinsic (resonance-based) \\
\bottomrule
\end{tabular}
\end{table*}

\subsection{Radiation hardness: qualitative comparison}
\label{sec:hardness}

The graphene-assisted interface is predicted to suppress leakage and retain a predicted sub-nanosecond response timescale up to $1~\mathrm{MGy}$ while preserving the sensitivity required for dosimetry (see Table~\ref{tab:sic}). This is achieved by tuning the capacitance ratio $\eta$ to balance radiation tolerance against dose-induced electrostatic modulation, which would otherwise degrade the resonant response. The predicted radiation tolerance is qualitatively consistent with recent experiments on graphene-optimized 4H-SiC detectors. Silicon carbide detectors are widely recognized for their inherent radiation tolerance~\cite{Nava2008}, and recent graphene-optimized 4H-SiC devices have further demonstrated stable operation under high-dose irradiation. In particular, graphene-optimized 4H-SiC p-i-n detectors have demonstrated stable charge-collection and fast transient response after proton irradiation~\cite{Jiang2026}, while a recent 
preprint reported only moderate degradation of charge-collection efficiency after exposure to 160-keV X-rays up to 1 MGy~\cite{Huang2026}. These results provide an experimental reference point for the MGy-class radiation-hardness target considered here. Unlike conventional detectors, the resonant transport mechanism offers inherent energy selectivity. Furthermore, the stronger carrier confinement in quantum-dot structures can also improve radiation tolerance relative to quantum-well or bulk counterparts~\cite{Huang2003,Aierken2013,Li2025}, which further supports the rationale for the DWELL architecture.

\section{FLASH Dose-Rate Discrimination}
\label{sec:flash}

\subsection{Three-population trap model}
\label{sec:threepop}

To reconcile the steady-state model with clinical dose rates, we introduce: deep traps ($\tau_{\mathrm{deep}}\gg$ experiment, dominate $Q_{\max}$ and $D_{\mathrm{sat}}$), slow traps ($\tau_{\mathrm{slow}}\sim$ seconds to hours), and fast traps ($\tau_{\mathrm{fast}}\sim 10~\mathrm{ms}$--$1~\mathrm{s}$). $Q_{\mathrm{tot}}=Q_{\mathrm{deep}}+Q_{\mathrm{slow}}+Q_{\mathrm{fast}}$, with $Q_{\mathrm{deep}}^{\max}\gg Q_{\mathrm{slow}}^{\max},Q_{\mathrm{fast}}^{\max}$. Only $Q_{\mathrm{deep}}$ contributes to steady-state dosimetry.

Fast traps follow
\begin{equation}
\frac{dQ_{\mathrm{fast}}}{dt} = (Q_{\mathrm{fast}}^{\max}-Q_{\mathrm{fast}})\frac{\dot{D}}{D_{\mathrm{fast}}} - \frac{Q_{\mathrm{fast}}}{\tau_{\mathrm{fast}}},
\label{eq:fast_trap}
\end{equation}
with the critical dose rate defined as
\begin{equation}
\dot{D}_{\mathrm{crit}} = \frac{D_{\mathrm{fast}}}{\tau_{\mathrm{fast}}}.
\label{eq:dcrit}
\end{equation}
Choosing $D_{\mathrm{fast}}\approx 50~\mathrm{Gy}$ and $\tau_{\mathrm{fast}}\approx 10~\mathrm{ms}$ yields $\dot{D}_{\mathrm{crit}}=5\times 10^3~\mathrm{Gy/s}$. Allowing $\tau_{\mathrm{fast}}$ to vary over $1~\mathrm{ms}$--$100~\mathrm{ms}$ gives a predicted critical range of $5\times 10^2$--$5\times 10^4~\mathrm{Gy/s}$, which overlaps a significant portion of the FLASH dose-rate regime ($\geq 40~\mathrm{Gy/s}$). The FLASH/UHDR regime is commonly associated with mean dose rates above approximately $40~\mathrm{Gy/s}$, although detector response depends strongly on dose per pulse, pulse structure, and beam time structure~\cite{FLASHdosimetry2022}.

\begin{figure}[!ht]
\centering
\includegraphics[width=0.49\textwidth]{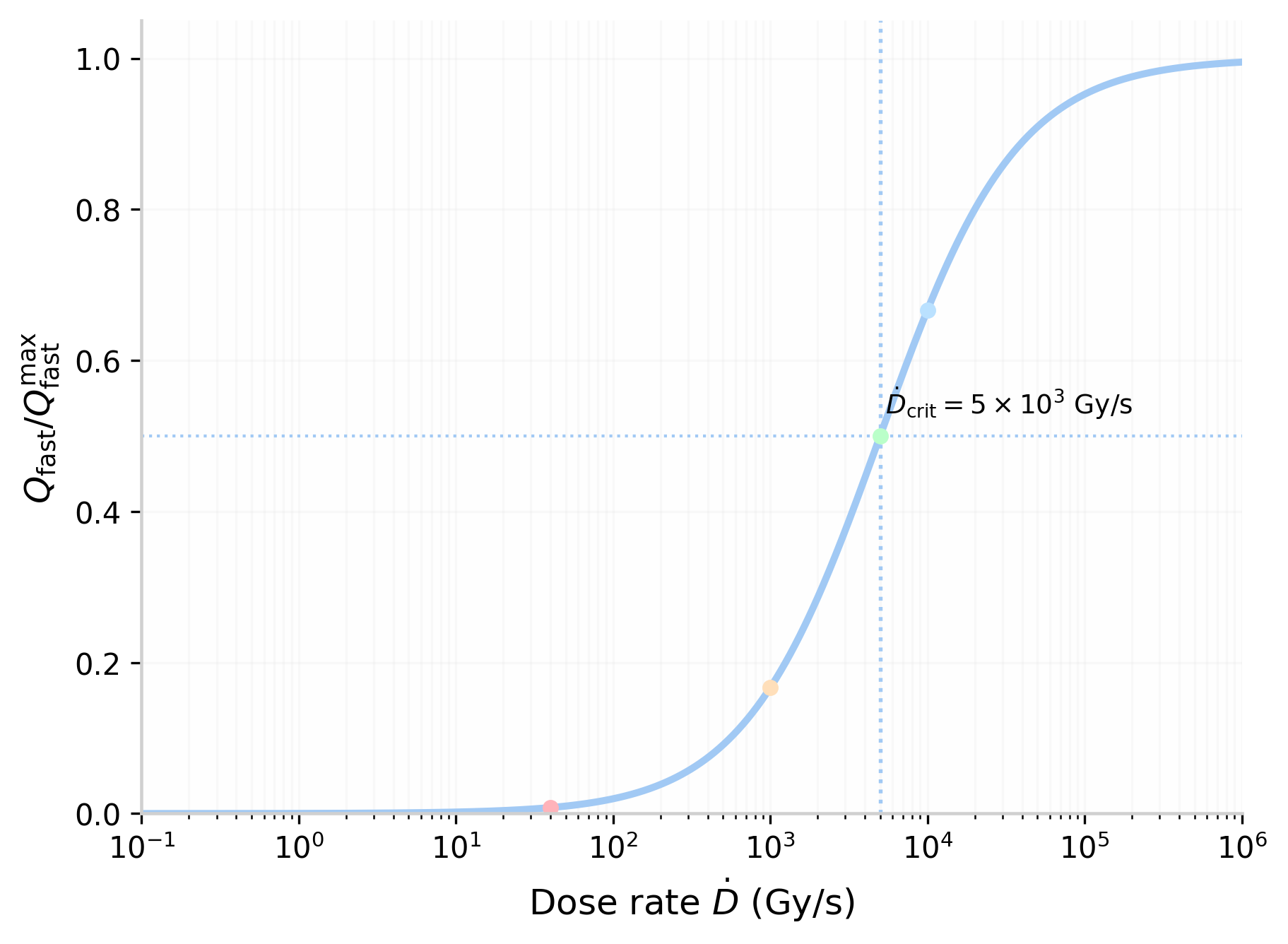}
\caption{\small Normalized fast-trap charge $Q_{\mathrm{fast}}/Q_{\mathrm{fast}}^{\max}$ versus dose rate $\dot{D}$. The vertical dotted line marks the critical dose rate $\dot{D}_{\mathrm{crit}}=5\times 10^3~\mathrm{Gy/s}$. Circles indicate representative FLASH-related rates.}
\label{fig:fast_trap}
\end{figure}

Figure~\ref{fig:fast_trap} illustrates the fast-trap saturation fraction
\begin{equation}
\frac{Q_{\mathrm{fast}}}{Q_{\mathrm{fast}}^{\max}} = \frac{\dot{D}}{\dot{D}+\dot{D}_{\mathrm{crit}}},
\label{eq:fast_fraction}
\end{equation}
with $\dot{D}_{\mathrm{crit}}=5\times 10^3~\mathrm{Gy/s}$. At conventional dose rates the fast-trap contribution is negligible. Within the broad FLASH/UHDR range, the fast-trap population becomes progressively populated, with substantial saturation occurring when $\dot{D}$ approaches or exceeds $\dot{D}_{\mathrm{crit}}=5\times10^3~\mathrm{Gy/s}$ for the nominal $\tau_{\mathrm{fast}}=10~\mathrm{ms}$. This provides a transient signal proportional to dose rate rather than accumulated dose.

\begin{figure}[!ht]
\centering
\includegraphics[width=0.49\textwidth]{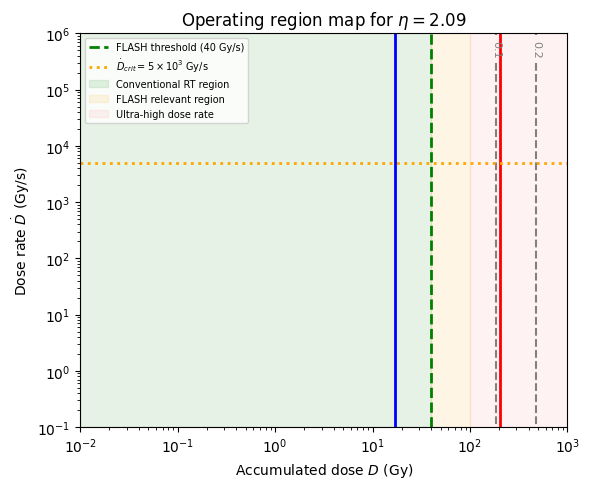}
\caption{\small Operating region map: normalized resonance shift $\Delta E(D,\eta)/\Delta E_{1,\max}^{(0)}$ as a function of accumulated dose $D$ and dose rate $\dot{D}$ for a representative screening parameter $\eta=2.09$. The contour lines denote the small-signal linearity threshold ($0.1\Gamma$) and saturation ($\Gamma$). The dashed vertical line marks the FLASH dose-rate threshold of $40~\mathrm{Gy/s}$.}
\label{fig:operating_regions}
\end{figure}

The trade-off between accumulated dose and dose rate is illustrated in Fig.~\ref{fig:operating_regions}, which shows the regions where the device remains in the linear regime versus saturation. This plot provides a compact visualization of the operational window for the proposed dosimeter.

\subsection{Toward radiation-type discrimination}
\label{sec:discrimination}

The ratio
\begin{equation}
R(D) = \frac{\beta_i^{\mathrm{eff}}}{\alpha_i^{\mathrm{eff}}}
\label{eq:ratio}
\end{equation}
is sensitive to the relative contributions of ionization and displacement processes. Slow recovery $\tau_{\mathrm{slow}}=[\sigma v_{\mathrm{th}}N_c]^{-1}\exp(E_a/k_BT)$ has species-dependent $E_a$ and $\sigma$, offering potential discriminative signatures.

\section{Conclusions}
\label{sec:conclusion}

Electrostatic interface engineering with graphene provides a physically motivated design framework for MR-DWELL dosimeters, enabling a tunable trade-off between dosimetric sensitivity and radiation tolerance. The capacitance ratio $\eta=C_Q/C_{\mathrm{geo}}$ provides a compact design parameter linking the h-BN thickness to the degree of electrostatic screening. At $\eta=2.09$, the model predicts a PVR of $\sim 10$ and $\sim 83\%$ peak-current retention after 1 MGy, with a maximum screened resonance shift of only $\sim 23$ meV. The second-derivative readout $d^2I/dV^2$ offers a practical experimental probe of radiation-induced resonance evolution.

The corrected formulation explicitly separates the independently calibrated low-dose sensitivity coefficient $\alpha_i^{\mathrm{eff}}$ from the saturation amplitude $\Delta E_{i,\max}$, thereby avoiding the previous mixing of two distinct dose scales. Their numerical scale difference is retained explicitly rather than conflated. A self-consistent NEGF--Poisson--trap framework is specified as a route toward future quantitative validation. The three-population trap model further extends the concept toward dose-rate discrimination in the FLASH/UHDR regime, while the estimated intrinsic graphene-interface RC time of $\sim 0.28$ ps indicates that the interface itself need not impose a fundamental sub-nanosecond bottleneck.

We emphasize that the present work is not a self-consistent NEGF calculation of the DWELL spectrum, but a reduced-order benchmark for the electrostatic screening physics. The literature-calibrated resonance energies ($E_{10}=82$ meV and $E_{20}=126$ meV) should not be interpreted as quantitatively reproduced by the self-consistent NEGF calculation for the identical geometry, which yields resonance groups near $\sim 44$ meV and $\sim 173$ meV~\cite{Dosymova2026d}. The screening framework based on $\eta$, $S$, and $\alpha_i^{\mathrm{eff}}$ remains useful because these quantities are determined by the capacitance ratio and the adopted trapped-charge saturation law rather than by the absolute resonance energies. Accordingly, the present results should be interpreted primarily as a quantitative benchmark for electrostatic screening trends, while absolute resonance positions and transport amplitudes require self-consistent validation.

This distinction between robust screening physics and geometry-specific absolute transport parameters defines the methodological scope of the present work. All quantitative predictions presented here are theoretical and require dedicated experimental validation to establish the practical viability of the proposed dosimeter concept.

\begin{acknowledgments}
This work was supported by the Tashkent Branch of the National Research Nuclear University MEPhI. The~authors thank colleagues for fruitful discussions. The~authors declare that no external funding was received for this theoretical study.
\end{acknowledgments}

\begin{appendix}
\section{Material Parameters}
\label{app:matparam}
For completeness and reproducibility, the material parameters used in the effective-mass Hamiltonian are summarized in Table~\ref{tab:matparams}.
\begin{table}[!ht]
\centering
\caption{Material parameters adopted for the effective-mass Hamiltonian (Refs.~\cite{Vurgaftman2001,Dosymova2026d}). The band-gap values are specified at 300~K; effective masses and dielectric constants are treated as temperature-independent parameters in the present model.}
\label{tab:matparams}
\begin{tabular}{lcccc}
\toprule
Parameter & GaAs & Al$_{0.3}$Ga$_{0.7}$As & In$_{0.15}$Ga$_{0.85}$As & InAs \\
\midrule
$m^*_e$ ($m_0$) & 0.067 & 0.092 & 0.060 & 0.023 \\
$m^*_h$ ($m_0$) & 0.34 & 0.38 & 0.35 & 0.41 \\
$E_g$ (eV) & 1.42 & 1.80 & 1.20 & 0.36 \\
$\epsilon_r$ & 12.9 & 12.2 & 13.2 & 15.1 \\
\bottomrule
\end{tabular}
\end{table}
\end{appendix}

\end{document}